\documentclass[prd,aps,preprint,tightenlines,showpacs,nofootinbib,superscriptaddress]{revtex4-1}
\usepackage{mathrsfs}
\usepackage{amsfonts}
\usepackage{amsmath}
\usepackage{array}
\usepackage{verbatim}
\usepackage{bm}
\usepackage{epsfig}
\usepackage{graphicx,color}
\usepackage{relsize}
\usepackage{lineno}
\usepackage{multirow}

\RequirePackage{xspace}

\begin{document}

\title{Production of vector mesons in pentaquark states resonance channel in
	$p$-$A$ ultraperipheral collisions}
\author{Ya-Ping Xie}
\email{xieyaping@impcas.ac.cn}
\affiliation{Institute of Modern Physics, Chinese Academy of Sciences,
Lanzhou 730000, China}
\author{Xiao-Yun Wang}
\email{xywang@lut.edu.cn (corresponding author)}
\affiliation{Department of physics, Lanzhou University of Technology,
Lanzhou 730050, China}
\author{Xurong Chen}
\email{xchen@impcas.ac.cn}
\affiliation{Institute of Modern Physics, Chinese Academy of
     		Sciences, Lanzhou 730000, China}
\affiliation{ University of Chinese
Academy of Sciences, Beijing 100049, China}
\affiliation{ Guangdong Provincial Key Laboratory of Nuclear Science, Institute of Quantum Matter, South China Normal University, Guangzhou 510006, China}
\begin{abstract}
	Ultraperipheral collisions (UPCs) of protons and nuclei are important for the
	study of the photoproduction of vector mesons and exotic states.
	The photoproduction of vector mesons in the pentaquark resonance channel in $p%
	\mbox{-}Au$ UPCs at the Relative Heavy Ion Collider (RHIC) and $p$-$Pb$ UPCs at
	the Large Hadron Collider (LHC) is investigated by
	employing STARlight package. The cross sections of vector mesons via
	pentaquark state resonances channel are obtained using effective Lagrangian
	method. The pseudo-rapidity and rapidity distributions of $J/\psi $ and $%
	\Upsilon(1S) $ are given for $p\mbox{-}Au$ UPCs at the RHIC and $p$-$Pb$ at the LHC.
	It is found that the RHIC is a better platform for discovering the pentaquark states than the LHC.
	Moreover, $P_{b}(11080)$ is easier to identify than $P_c(4312)$ because the background of $\Upsilon (1S)$ is weaker
	than $J/\psi$ in the $t$-channel at the RHIC.
\end{abstract}

\maketitle



\section{introduction}

Ultraperipheral collisions (UPCs) are important tools for investigating
photoproduction at high energies\cite{Baur:2001jj,Bertulani:2005ru,Baltz:2007kq}. UPCs
can probe $\gamma$-$\gamma$ and $\gamma$-$h$ interactions via vector
meson production and dijet production\cite{Klein:2017vua}. UPCs have been studied
at  the Large Hadron Collider (LHC) and Relative Heavy Ion Collider (RHIC) for
photoproduction of several vector mesons, such as $\rho$, $\omega$, $\phi$, $%
J/\psi$. The photoproduction of exotic particles by UPCs have also been investigated
\cite{Klein:2019avl,Goncalves:2019vvo}.

In UPCs, the impact parameter of two hadrons is larger than the sum radius of two hadrons.
The direct strong interaction between two hadrons is thereby suppressed due to the large distance.
However, the accelerated hadrons are surrounded by a cloud of photons with virtuality $Q^2<(\hbar c/R_A)^2$ .
if $A>16$ and $Q^2$ is less than $($60 $\mathrm{MeV})^2$, a photon is almost a real photon in UPCs\cite{Baltz:2007kq}.
Therefore, electromagnetic
interaction is important in UPCs since it is a long range interaction.
The photon-hadron interaction can also be investigated in UPCs. Since the direct strong reaction between
hadrons is suppressed in UPCs, the background of UPCs is cleaner than the non-UPCs in hadron-hadron collisions.

Recently, several narrow pentaquark states named $%
P_{c}(4312),P_{c}(4440),P_{c}(4457)$ have been observed in $\Lambda
_{b}\rightarrow J/\psi pK$ by the LHCb Collaboration \cite{Aaij:2015tga,Aaij:2019vzc}, which is an important
progress in the search for exotic hadrons.
Many theoretical models have been proposed to study the internal natural and production of pentaquark states $P_c$
since there discovery \cite{Liu:2019tjn,Chen:2016qju,He:2019ify,Chen:2019asm,Ali:2019npk,Xiao:2019mst,Fernandez-Ramirez:2019koa,Guo:2019fdo,Zhu:2019iwm,Voloshin:2019aut,Wu:2019rog,Albaladejo:2020tzt,Wang:2019dsi,Wang:2019krd}.

Since the intermediate particles
in the above reaction process satisfy the on-shell condition, the
contribution of triangular singularities in the $\Lambda _{b}\rightarrow
J/\psi pK$ reaction cannot be ignored \cite{guo2020}. This means that it is
currently difficult to determine whether these $P_{c}$ states are
genuine state. However, one can observe and study the $P_{c}$ state via
other scattering processes, such as photoproduction processes, thereby
effectively avoiding the influence of triangular singularities in order to confirm
whether $P_{c}$ state is a genuine state. In Refs. \cite%
{Wang:2019krd,Wang:2019dsi}, combined with the latest experimental results,
an in-depth study of $P_{c}$ state production via $\gamma p$ or $\pi
^{-}p$ was carried out. Subsequently, the GlueX Collaboration reported their
first measurement of the $\gamma p\rightarrow J/\psi p$ process~\cite{Ali:2019lzf}. Although
the GlueX group did not find the photoproduction of pentaquark states with the
present precision~\cite{Ali:2019lzf}, their data suggested a meaningful upper limit of
production cross sections and hence a model dependent upper limit of
branching ratios $\mathcal{B}(P_{c}\rightarrow J/\psi p)$ of a small percent-age at most.
The size of the branching ratio of $%
P_{c}\rightarrow J/\psi p$ suggested by experiment is largely consistent
with the results in Ref. \cite{Wang:2019krd}. In Ref. \cite{Wang:2019zaw}, based
on the previous predictions of the mass and width of the hidden bottom
pentaquark $P_{b}$, a systematic study of the photoproduction of the $P_{b}$
state was conducted. These photoproduction results of $P_{c}$ and $P_{b}$
are very important foundations for studying the production of pentaquark
states via  UPCs.

STARlight is a Monte-Carlo package for vector meson production simulation in UPCs\cite{Klein:2016yzr}.
It is widely used
in $AA$ and $pA$ UPCs at the LHC and RHIC. The cross section calculation of photon-proton
to vector mesons is needed in STARlight package, and the total cross section
of vector meson in UPCs can be obtained by multiplying the photon flux. We use the photon-induced cross section of
pentaquark states of $P_c$ and $P_b$ and implement the cross sections in STARlight.
In this way, we can obtain the total cross sections of vector meson in UPCs and simulate the distributions.
The output of STARlight is four momentum of the final states. Using the four momentum of the final states,
one can obtain the rapidity and pseudo-rapidity distributions of the vector mesons.
We apply STARlight package to simulate vector mesons
in the pentaquark resonance $s$-channel and pomeron exchange $t$-channel. The
pseudo-rapidity and rapidity distributions are presented in this paper.

The aim of this paper is to study the production of hidden charm/bottom
pentaquark states in UPCs. The relevant results can provide an important
theoretical basis for finding $P_{c}$ and $P_{b}$ states via future UPCs
experiments. This paper is organized as follows. The theoretical
framework is presented in Section~\ref{sec:framework}. The numerical results are
given in Section.~\ref{sec:numerical}, and a summary concludes the paper  in Section.~\ref%
{sec:conclusion}.

\section{Theoretical Framework}

\label{sec:framework} 
\begin{figure}[h]
	\centering
	\includegraphics[width=0.8\textwidth]{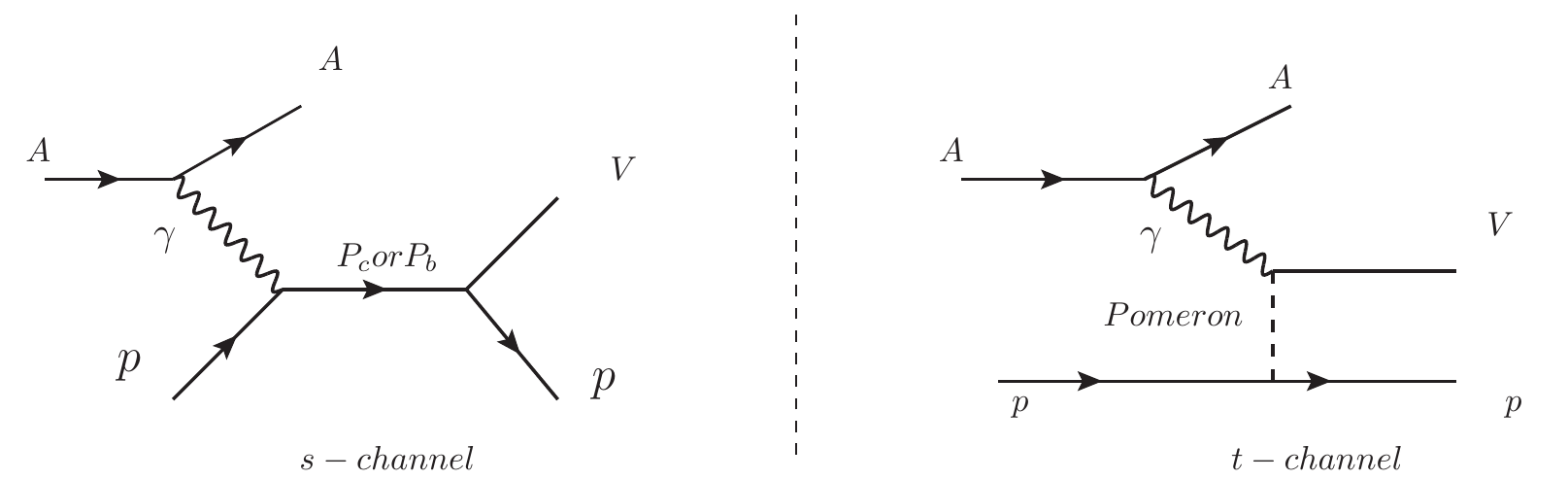}
	\caption{ The processes of vector mesons production in $p\mbox{-}A$ UPCs via
		$s$-channel (left graph) and $t$-channel (right graph). }
	\label{fig01}
\end{figure}

In UPCs, a real photon emitted from one nucleus can interact
with a nucleus from other direction. Because
the photon number is proportional to the charge number of the hadrons, the photon flux emitted
from proton can be
neglected when compared to the photon flux emitted from nucleus in $p\mbox{-}A$ UPCs, for example, those of gold and Lead.
Thus, in $p\mbox{-}A$ UPCs, we can neglect the photons emitted from proton. We only consider
the photons from nucleus. Diagrams of the $p$-$A$ processes in the $s$-channel and $t$-channel schemes
are shown in Fig.~\ref{fig01}. In the $s$-channel, the vector mesons are
produced via pentaquark state resonance. In the $t$-channel, the photon interacts with proton via pomeron exchange
and produces vector mesons.
As the cross section of the $t$-channel is dominant in photon-proton interaction, the cross section of
the $t$-channel be viewed as a background of the pentaquark resonance in discovering pentaquark states processes.
Usually, for an electromagnetic scattering process, the scattering amplitude should meet the requirements of gauge invariance.
In the process shown in Fig,~\ref{fig01}, the $s$-channel amplitude $\mathcal{M}$ satisfies the relation $k\cdot \mathcal{M}=0$,
when $k$ is the photon momentum. Moreover, the cross section of the $t$-channel can be calculated by a parameterized pomeron
model rather than by constructing amplitude. Therefore, for the current scattering process, we posit that the scattering amplitude
roughly satisfies gauge invariance.

Because the impact parameter of two hadrons is large in UPCs, the production of pentaquark states via
$\pi$ meson exchanging contributions in hadrons can be neglected in UPCs since $\pi$ meson exchanging
interaction is a short range interaction. We thus only consider the photon-induced pentaquark states production
in UPCs as depicted  in Fig.~\ref{fig01}.

The cross section of  the vector mesons in UPCs is computed by integrating the
photon flux and photon-proton cross section. The photon flux can be
obtained in QED calculations. The photon-proton cross section for vector
mesons can be calculate in several models. In $p\mbox{-}A$ UPCs, the cross section of the $%
pA\rightarrow pAV$ is given as follows \cite{Klein:2016yzr}
\begin{equation}
\sigma (pA\rightarrow pAV)=\int dk\frac{dN_{\gamma }(k)}{dk}\sigma _{\gamma
	p\rightarrow pV}(W).\label{eqcs}
\end{equation}
where $k$ is the momentum of the photon emitted from nucleus, and $W$ is the
center of mass (c.m.) energy of the photon and proton system. The relationship of rapidity
of vector meson $\mathrm{y}$ and $k$ ,$W$ are $k=M_V/2\exp(\mathrm{y})$, $W=(2k\sqrt{s})^{1/2}$. Using these equations,
we obtain the rapidity
distributions of vector meson in UPCs\cite{Xie:2018rog}.
\begin{eqnarray}
\frac{d\sigma}{d\mathrm{y}}=k\frac{dN_{\gamma }(k)}{dk}\sigma _{\gamma
	p\rightarrow pV}(W).
\end{eqnarray}
The photon
emitted from nucleus is presented as \cite{Klein:1999qj}
\begin{equation}
\frac{dN_{\gamma }(k)}{dk}=\frac{2Z^{2}\alpha }{\pi k}\big(XK_{0}(X)K_{1}(X)-%
\frac{X^{2}}{2}[K_{1}^{2}(X)-K_{0}^{2}(X)]\big).\label{phflux}
\end{equation}
where $X=b_{min}k/\gamma_L $, $b_{min}=R_{A}+R_{p}$ is sum of radius of proton
and nucleus. $\gamma_L $ is the Lorentz boost factor, and it is obtained by $\gamma_L=\sqrt{s}/2m_p$. $K_{0}(x)$ and $K_{1}(x)$
are the Bessel functions. By employing the cross section of vector meson in $p\mbox{-}A$
UPCs, we obtain the total cross section in $p\mbox{-}Au$ UPCs at RHIC experiments.
From the Eq.(\ref{phflux}), it can be seen that the photon flux is proportional to the charge
number of hadron. For example, the photon flux of gold is $Z^2 \approx 6200$ larger than the photon flux of
proton $Z^2=1$. We can therefore neglect the contributions of photon from proton in $p$-$A$ UPCs.

In Ref. \cite{Wang:2019krd,Wang:2019zaw}, the cross sections of $\gamma
p\rightarrow P_{c}\rightarrow J/\psi p$ and $\gamma p\rightarrow
P_{b}\rightarrow \Upsilon (1S) p$ via the $s$-channel were calculated based on the
effective Lagrangian method and vector-meson-dominance (VMD) model. In the
photoproduction calculation, the branching ratio of $P_{c}(4312)\rightarrow
J/\psi p$ was taken as 3\%, and the decay width of $P_{b}(11080)\rightarrow
\Upsilon (1S) p$ was taken as 0.38 MeV predicted in Ref.~\cite{Gutsche:2019mkg}. The
numerical result shown that the average value of the cross section from the $%
P_{c}(4312)$ or $P_{b}(11080)$ produced in photon-proton scattering reaches
at least 0.1 nb with a bin of 0.1 GeV. In this work, we employ the results for $%
P_{c}/P_{b}$ photoproduction from Refs. \cite{Wang:2019krd,Wang:2019zaw}
to calculate the production of $P_{c}/P_{b}$ via UPCs.

For the contribution of $t$-channel Pomeron exchange, the cross section of $%
\gamma p\to V p$ is given as \cite{Klein:2016yzr},
\begin{eqnarray}
\sigma^t_{\gamma p\to V p}(W)=\sigma_p\cdot\Big(1-\frac{(m_p+m_{V})^2}{W^2}%
\Big)\cdot W^\epsilon,\label{tccss}
\end{eqnarray}
with $\sigma_p$=4.06 nb and $\epsilon=0.65$ for $J/\psi$ and $\sigma_p$=6.4
pb and $\epsilon=0.74$ for $\Upsilon (1S)$, which are determined by the
experimental data of processes $\gamma p\to V p$.

By employing the cross sections in the $s$-channel and $t$-channel, we can obtain the
vector meson cross sections in $p\mbox{-}A$ UPCs. With the Monte-Carlo package
STARlight, we can simulate the vector meson production processes and obtain the
four momentum of final states. We then obtain the spectrum of vector mesons in two channels.

\section{Numerical result}

\label{sec:numerical} In this study, the cross sections of the vector mesons $%
J/\psi$ and $\Upsilon (1S)$ in the $t$-channel can be calculated using Eqs.(\ref%
{eqcs})-(\ref{tccss}). We use the same calculation progress of
vector mesons in the $s$-channel as \cite{Wang:2019krd,Wang:2019zaw}. STARlight
package is employed to simulate the vector mesons through the $t$-channel and $%
s$-channel in $p\mbox{-}Au$ UPCs at RHIC $p$-$Pb$ UPCs at the LHC. Vector meson distributions
are presented for the RHIC and LHC.

First, we calculate the $J/\psi $ and $\Upsilon (1S) $ cross sections in the $s
$-channel and $t$-channel in $p$-$Au$ and $p$-$Pb$ UPCs. The cross sections are listed in Table~\ref%
{table01} and Table~\ref{table02}, where the masses and decay widths are also listed. The event
numbers are also included, they can be applied to
estimate event number for one-year running for the RHIC and LHC.

\begin{table}[h]
	\begin{tabular}{|c|c|c||c|c|c|}
		\hline\hline
		{Resonance} &  & {Properties~\cite{Aaij:2019vzc,Gutsche:2019mkg}} &  & $s$-channel &
		$t$-channel \\ \cline{1-3}\cline{4-6}
		\multirow{2}{*}{ $P_c(4312)$} & Mass & $4311.9\pm 0.7_{-0.6}^{+6.8}$ MeV & $%
		J/\psi $ cross section & 1.8 nb & 2.2 $\mu $b \\ \cline{2-6}
		& Decay width & $9.8\pm 2.7_{-4.5}^{+3.7}$ MeV & Event Number & 8.1 K & 9.9 M
		\\ \hline\cline{1-3}\cline{4-6}
		\multirow{2}{*}{ $P_b(11080)$} & Mass & 11080 MeV & $\Upsilon (1S)$ cross
		section & 0.10 nb & 1.2 nb \\ \cline{2-6}
		& Decay width & 1.58 MeV & Event Number & 0.45 K & 5.4 K \\
		\hline\cline{2-6}\hline\hline
	\end{tabular}%
	\caption{Cross sections $J/\protect\psi $ and $\Upsilon (1S)$ in $%
		pAu\rightarrow pVAu$ in $s$-channel and $t$-channel. The collisions energy is $\sqrt{s}$= 200 GeV and the luminosity of the $p\mbox{-}Au$ is
		4.5 $\mathrm{pb}^{-1} $\protect\cite{Tanabashi:2018oca} }
	\label{table01}
\end{table}
From Table.~\ref{table01} and Table.~\ref{table02}, it can be seen that the cross sections of LHC are much larger than those
in the RHIC in the $t$-channel, and in the $s$-channel, the cross sections of the LHC are several times larger than
those of the RHIC. The reason
is that the cross section of the $t$-channel is dependent on the $W$ region. The $W$ region at the LHC is
boarder than that at the RHIC. However, in the $s$-channel, the $W$ regions are of the same size at the RHIC
and LHC.

\begin{figure}[h]
	\centering
	\includegraphics[width=0.45\textwidth]{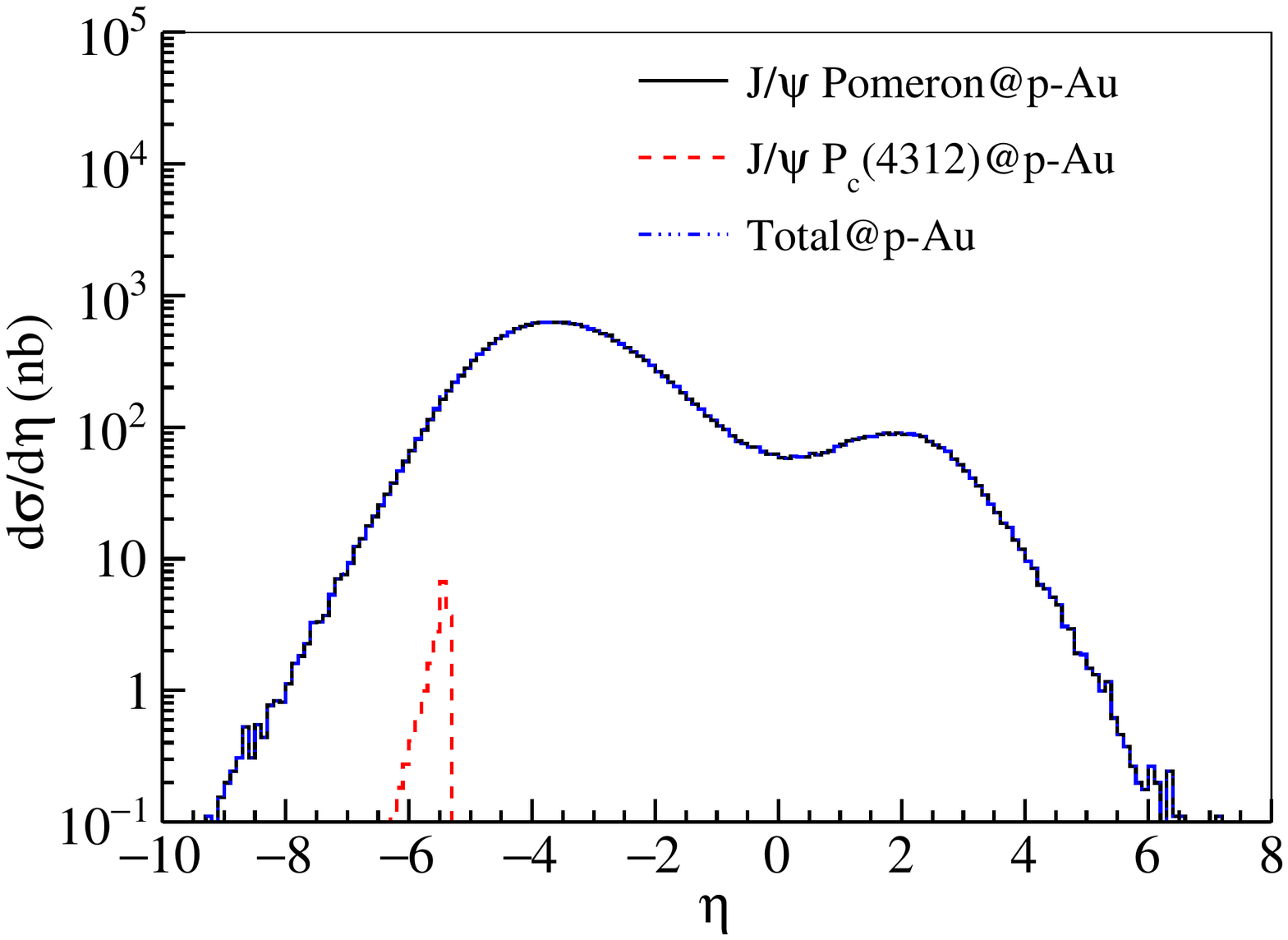} %
	\includegraphics[width=0.45\textwidth]{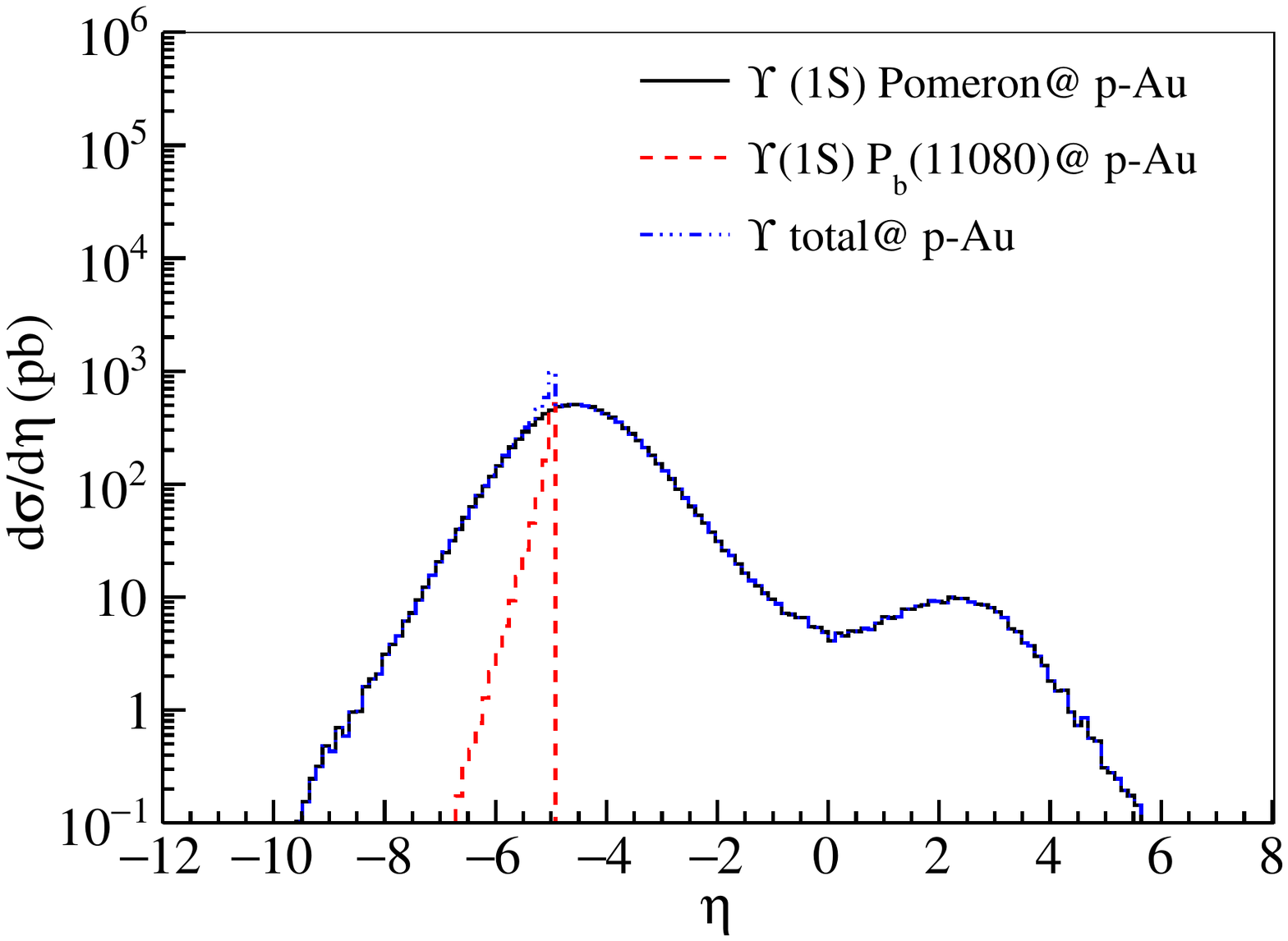}
	\caption{(Color online) Pseudo-rapidity distributions of $J/\protect\psi$
		and $\Upsilon (1S)$ produced from pomeron exchange $t$-channel (black solid
		curve) and pentaquark resonance $s$-channel (red dashed curve) $p\mbox{-}Au$
		UPCs at $\protect\sqrt{s}=200$ GeV at RHIC. The direction of gold beam is the positive pseudo-rapidity direction. }
	\label{fig02}
\end{figure}

\begin{table}[h]
	\begin{tabular}{|c|c|c||c|c|c|}
		\hline\hline
		{Resonance} &  & {Properties~\cite{Aaij:2019vzc,Gutsche:2019mkg}} &  & $s$-channel &
		$t$-channel \\ \cline{1-3}\cline{4-6}
		\multirow{2}{*}{ $P_c(4312)$} & Mass & $4311.9\pm 0.7_{-0.6}^{+6.8}$ MeV & $%
		J/\psi $ cross section & 7.4 nb & 0.10 mb \\ \cline{2-6}
		& Decay width & $9.8\pm 2.7_{-4.5}^{+3.7}$ MeV & Event Number & 7.4 K & 0.10 G
		\\ \hline\cline{1-3}\cline{4-6}
		\multirow{2}{*}{ $P_b(11080)$} & Mass & 11080 MeV & $\Upsilon (1S)$ cross
		section & 0.78 nb & 0.22 $\mu$b \\ \cline{2-6}
		& Decay width & 1.58 MeV & Event Number & 0.78 K & 0.22 M \\
		\hline\cline{2-6}\hline\hline
	\end{tabular}%
	\caption{Cross sections $J/\protect\psi $ and $\Upsilon (1S)$ in $%
		pPb\rightarrow pVPb$ in $s$-channel and $t$-channel.  The collision energy is $\sqrt{s}$ = 8.8 TeV and the luminosity of the $p$-$Pb$ is
		1 $\mathrm{pb}^{-1} $\protect\cite{Coelho:2020lyd} }
	\label{table02}
\end{table}
Second, the pseudo-rapidity distributions corresponding to the angle
distributions of two vector mesons of $p$-$Au$ UPCs at the RHIC are illustrated in Fig.~\ref{fig02}. It
can be seen that $J/\psi $ and $\Upsilon (1S)$ in the $s$-channel are totally
covered by the $t$-channel distributions. This is because the cross section of the $t$%
-channel is much larger than the $J/\psi $ cross section through pentaquark
exchange $s$-channel. The sum of the $s$-channel and $t$-channel is the same as the $%
t$-channel. As a result, it is difficult to identity the pentaquark signal
through $J/\psi+p$ invariant mass spectrum in pseudo-rapidity distributions.
In contrast, in $\Upsilon (1S)$ production, the $s$-channel signal is
significant for identifying the pentaquark states $P_b(11080)$  through $%
\Upsilon (1S)+p$ invariant mass spectrum in pseudo-rapidity distributions.

\begin{figure}[h]
	\centering
	\includegraphics[width=0.45\textwidth]{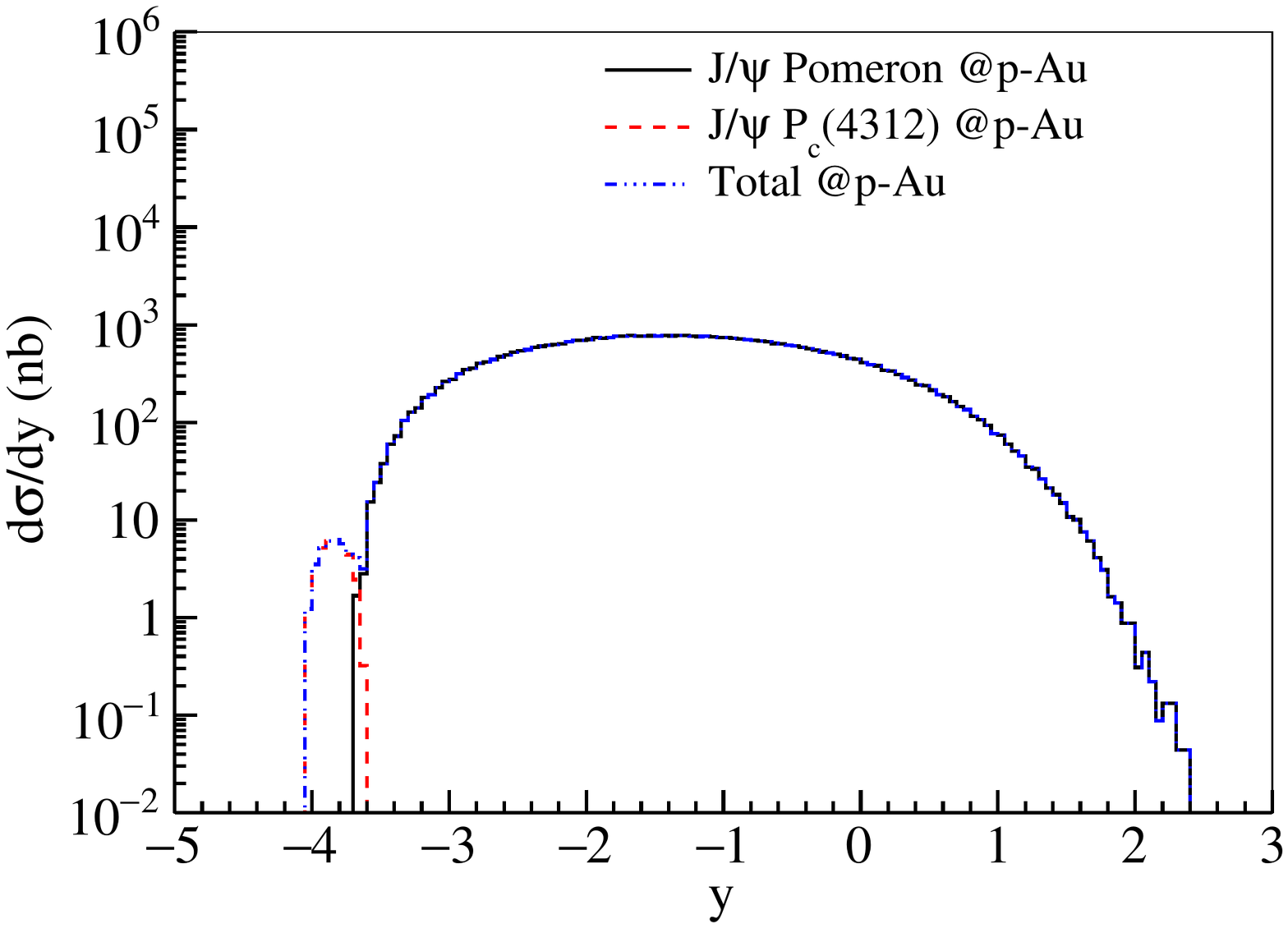} %
	\includegraphics[width=0.45\textwidth]{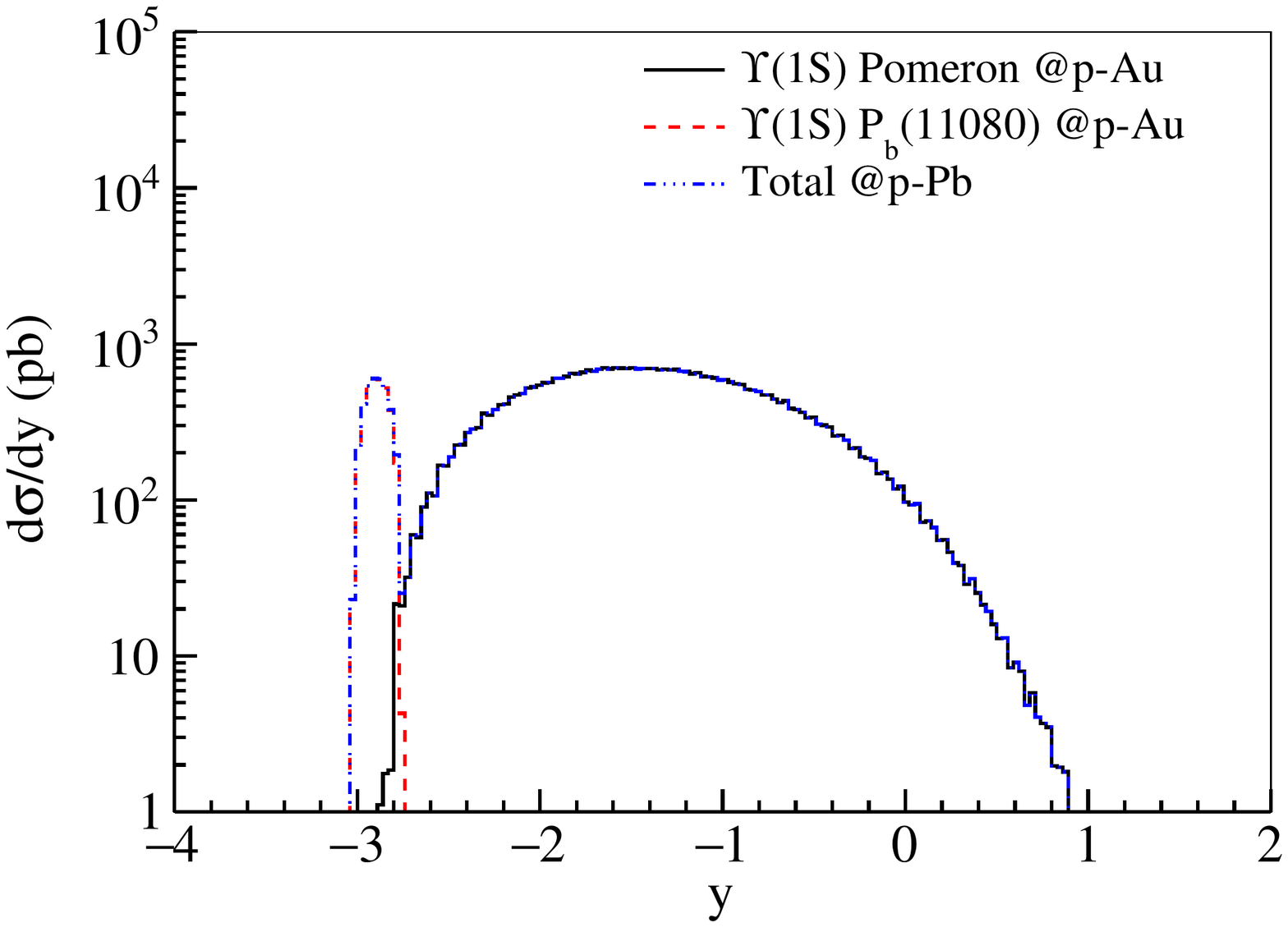}
	\caption{(Color online) Rapidity distributions of $J/\protect\psi$ and $%
		\Upsilon (1S)$ produced from pomeron exchange $t$-channel (black solid
		curve) and pentaquark resonance $s$-channel (red dashed curve) $p\mbox{-}Au$
		UPCs at $\protect\sqrt{s}=200$ GeV at RHIC. The direction of gold beam is the positive rapidity direction}
	\label{fig03}
\end{figure}

We also provide the rapidity distributions of two vector mesons in $p%
\mbox{-}Au$ UPCs in Fig. \ref{fig03}. We can see that in the rapidity space,
the vector mesons in the $s$-channel are not totally covered by the vector
mesons distributions in the $t$-channel. These results differ from the
rapidity distributions in Ref.\cite{Goncalves:2019vvo}. The rapidity
distributions of $J/\psi$  in the $s$-channel are totally
covered by the $t$-channel $J/\psi$ rapidity distributions in Ref. \cite%
{Goncalves:2019vvo}.  Moreover, it can be seen that the rapidity
distributions differ from the pseudo-rapidity distributions because
the energies in the same pseudo-rapidity region are different in Fig.~\ref%
{fig02} and Fig.~\ref{fig03}.
\begin{figure}[h]
	\centering
	\includegraphics[width=0.45\textwidth]{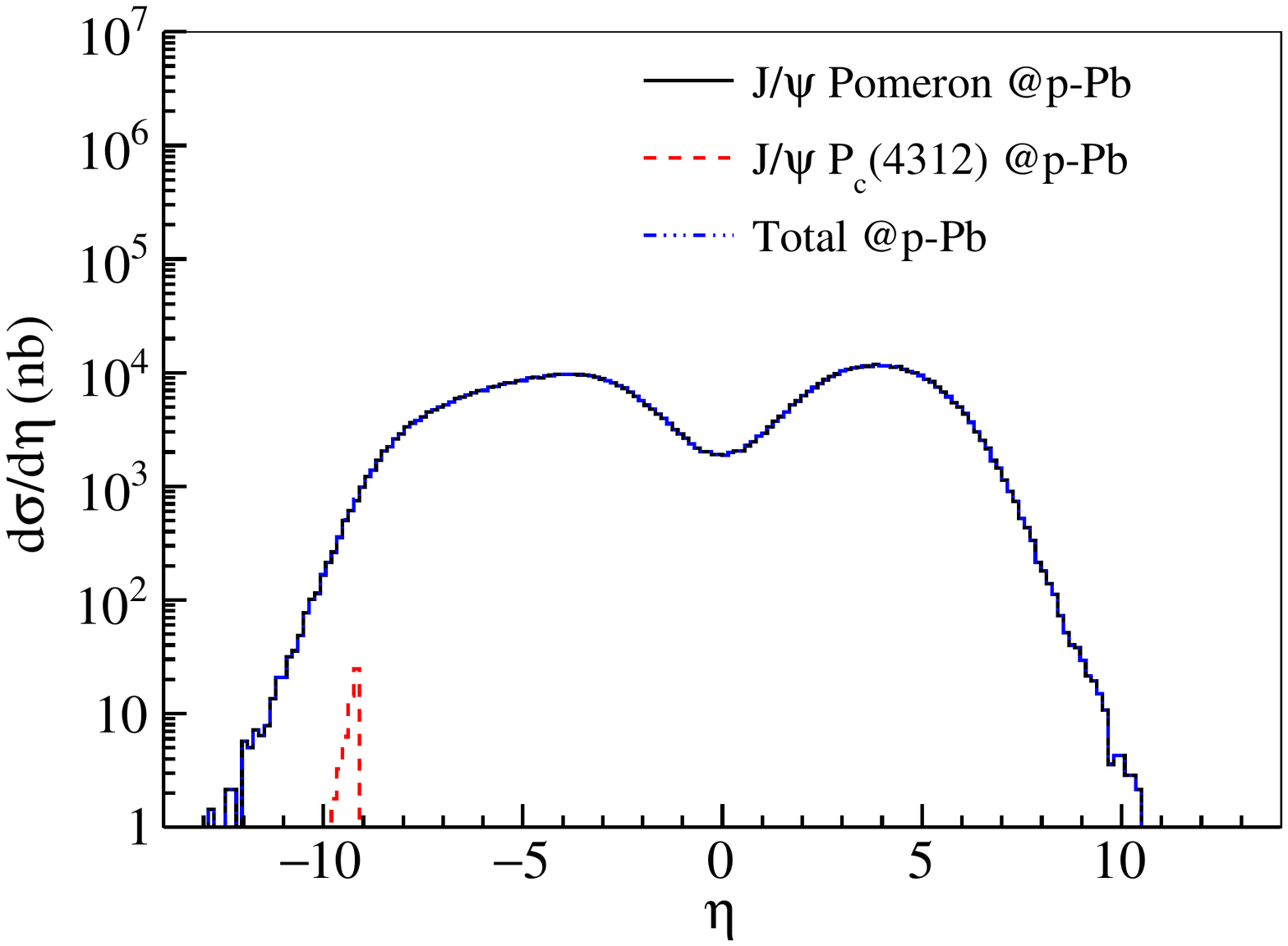} %
	\includegraphics[width=0.45\textwidth]{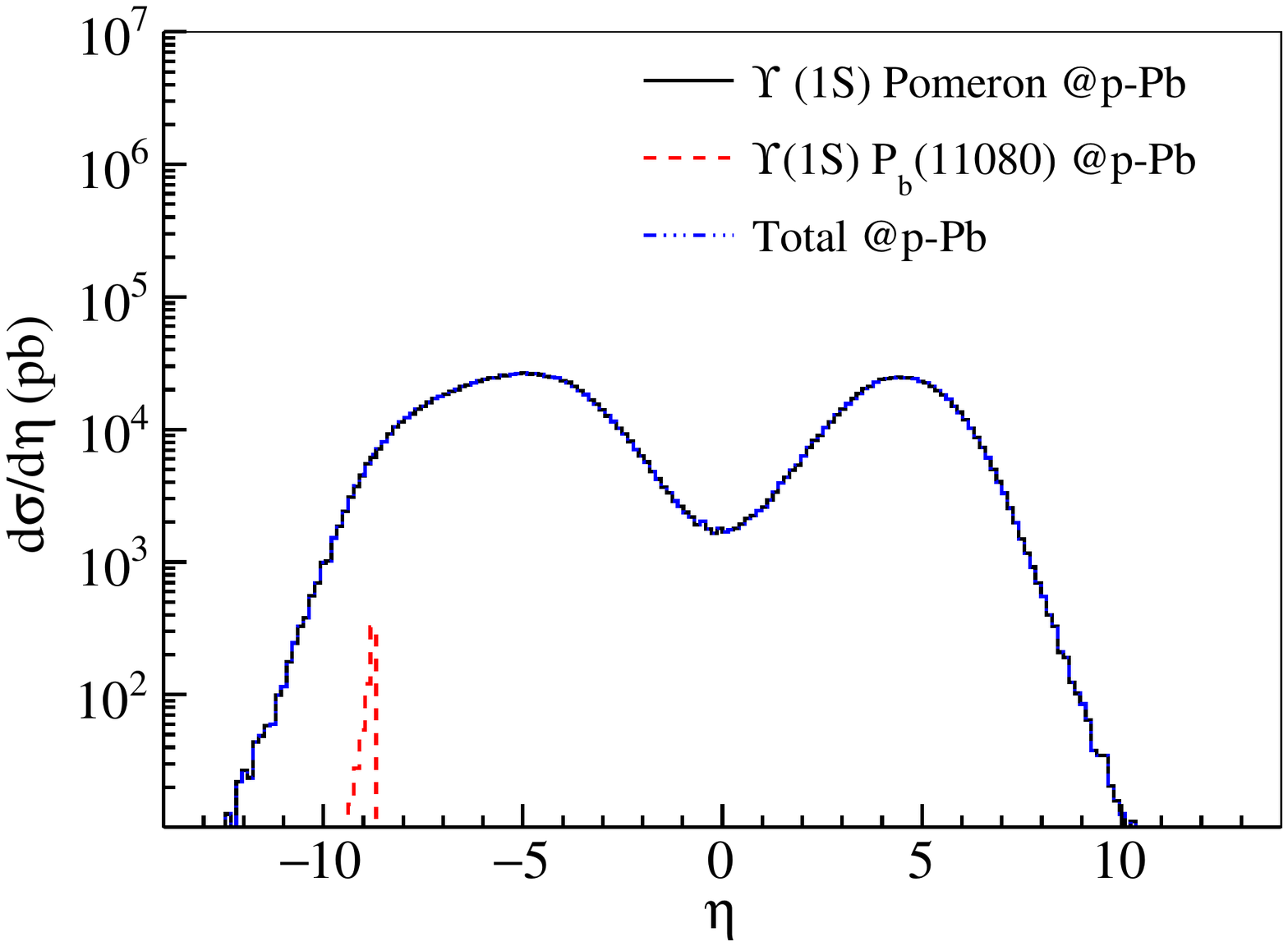}
	\caption{(Color online) Pseudo-rapidity distributions of $J/\protect\psi$
		and $\Upsilon (1S)$ produced from pomeron exchange $t$-channel (black solid
		curve) and pentaquark resonance $s$-channel (red dashed curve) $p\mbox{-}Pb$
		UPCs at $\protect\sqrt{s}=8.8$ TeV at the LHC. The direction of lead beam is the positive pseudo-rapidity direction. }
	\label{fig04}
\end{figure}
\begin{figure}[h]
	\centering
	\includegraphics[width=0.45\textwidth]{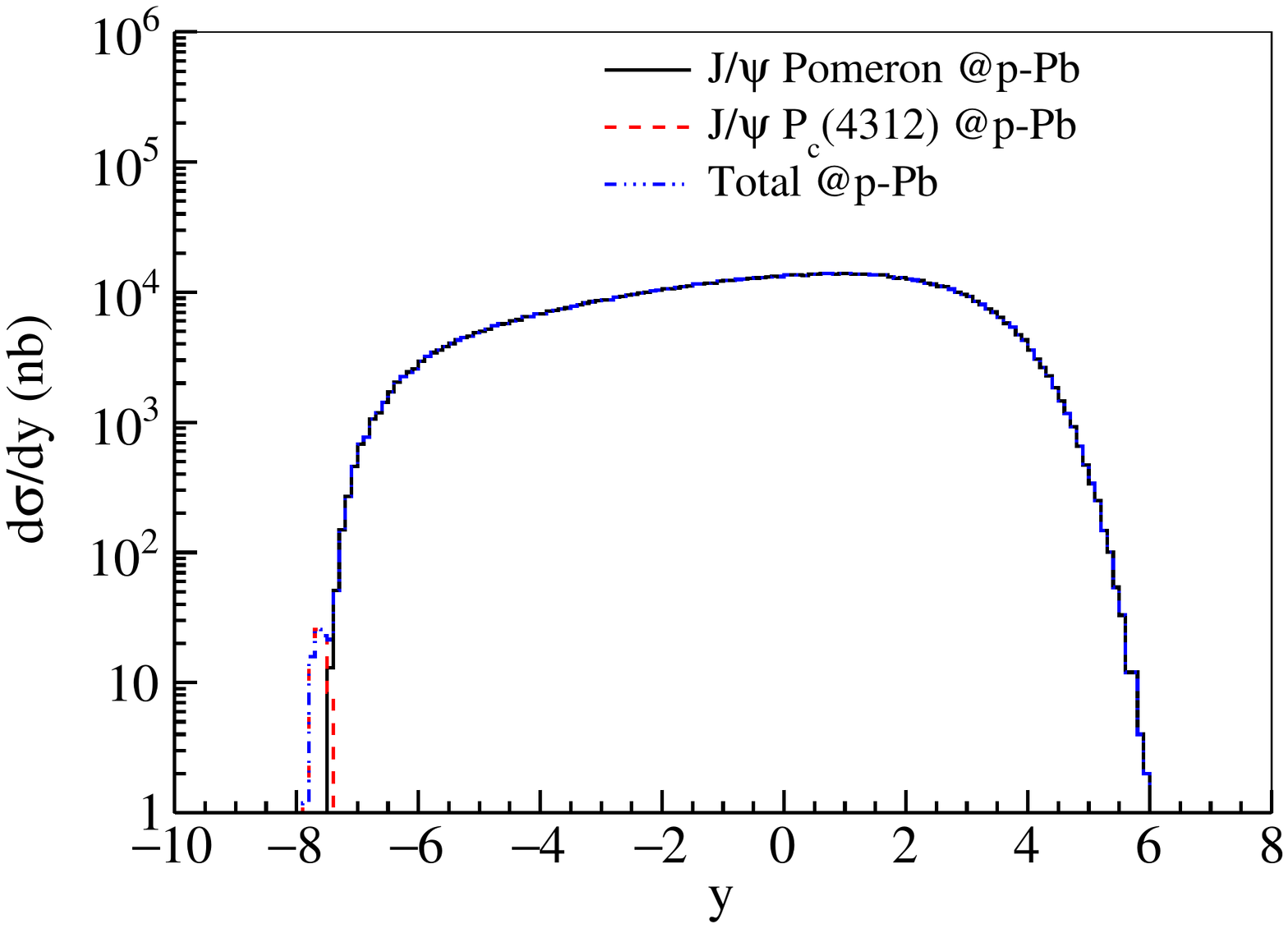} %
	\includegraphics[width=0.45\textwidth]{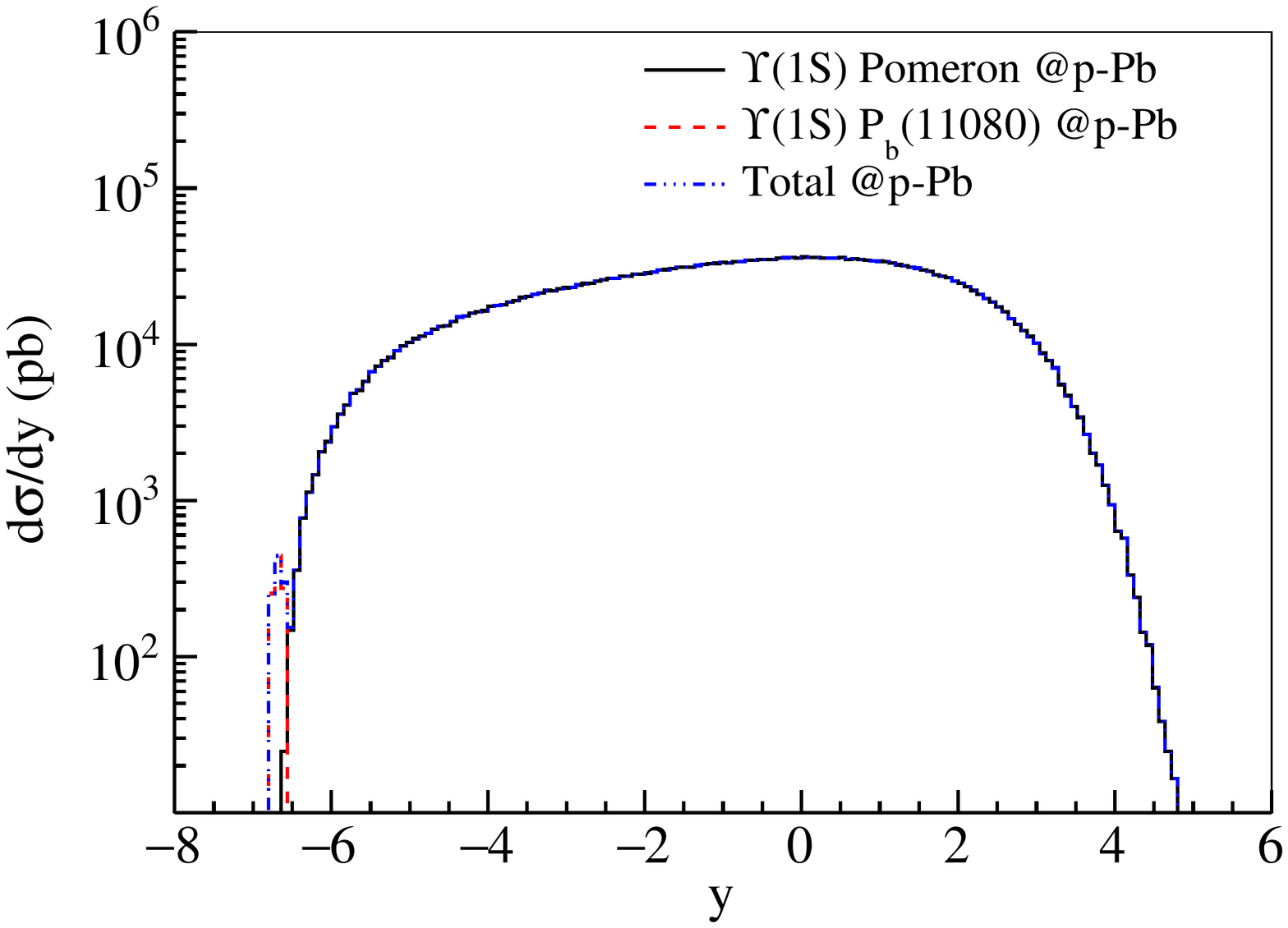}
	\caption{(Color online) Rapidity distributions of $J/\protect\psi$
		and $\Upsilon (1S)$ produced from pomeron exchange $t$-channel (black solid
		curve) and pentaquark resonance $s$-channel (red dashed curve) $p\mbox{-}Pb$
		UPCs at $\protect\sqrt{s}=8.8$ TeV at the LHC. The direction of lead beam is the positive rapidity direction. }
	\label{fig05}
\end{figure}

Furthermore, we present the production of $J/\psi$ and $\Upsilon(1S)$ via the $s$-channel and $t$-channel
for $p$-$Pb$ UPCs at the LHC. The predictions are shown in Fig.~\ref{fig04} and Fig.~\ref{fig05}.
It is evident that the distributions of the $t$-channel at the LHC are larger than those at the RHIC.
The background at the LHC is thus stronger than the background at the RHIC,
making it more difficult to identify the pentaquark states at the LHC than at the RHIC.\\
\indent Finally, the production of $J/\psi$ and $\Upsilon (1S)$ is presented for $p%
\mbox{-}Au$ UPCs at the RHIC and $p$-$Pb$ at the LHC. The cross sections in the $s$-channel and $t$-channel
, which can be used to predict the total detected event numbers at the
RHIC and LHC, are listed in Table.~\ref{table01} and \ref{table02}. The pseudo-rapidity and rapidity distributions of $J/\psi$ and $%
\Upsilon (1S)$ in the two channels are illustrated for $p\mbox{-}Au$ at the RHIC and $p$-$Pb$ at the LHC.
We can see that the pseudo-rapidity distributions and rapidity distributions
are different. It is concluded that RHIC is a better platform for the discovery of the pentaquark state
than the LHC, and that $P_{b}(11080)$ is a better candidate than $P_c(4312)$ at the RHIC because the background of
$\Upsilon (1S)$ is weaker than that of $J/\psi$ at the RHIC.

\section{conclusion}

\label{sec:conclusion} In this paper, we study the production of the vector
mesons  $J/\psi $ and $\Upsilon (1S)$ in pentaquark
resonance channel in $p\mbox{-}Au$ UPCs at the RHIC and $p$-$Pb$ UPCs at the LHC.
The cross sections pf $%
\gamma p\to Vp$ are computed via two channels. The vector meson and proton
production in the $s$-channel can be used to reconstruct pentaquark states. The
vector meson and proton production in the $t$-channel can be viewed as a background
of  the $s$-channel production.  The cross sections of $\gamma p\to J/\psi p$ and
$\gamma p\to \Upsilon (1S) p$ in the $s$-channel are calculated via effective
Lagrangian method. We apply STARlight package to simulate the vector meson
production in $p\mbox{-}Au$ UPCs at the RHIC and $p$-$Pb$ UPCs at the LHC.
We obtain several
distributions for $J/\psi$ and $\Upsilon (1S)$. The pseudo-rapidity and
rapidity distributions  of $J/\psi$ and $\Upsilon (1S)$ are illustrated in
this work. From these distributions,
We find that the background of the LHC is stronger than the that of the RHIC due to the
high collision energy. The RHIC is a better platform for identifying the pentaquark states.
Moreover, it can
concluded that the $P_{c}(4312)$ state is difficult to identify through $%
J/\psi+p$ invariant mass spectrum in $p\mbox{-}Au$ UPCs at $\sqrt{s}=200$
GeV at the RHIC. However, $P_{b}(11080)$ may be discovered in $\Upsilon
(1S)+p$ invariant mass spectrum at $p\mbox{-}Au$ UPCs at $\sqrt{s}=200$ GeV
at the RHIC, although a small energy bin width is necessary.
Consequently, it is important to detect $\Upsilon (1S)$
production in $p\mbox{-}Au$ UPCs at the RHIC, to aid in the discovery the
pentaquark $P_{b}(11080)$ .
\section*{Acknowledgment}

The work is supported by the National Natural Science Foundation of China
(Grant Nos. 11705076, 11975278, 11405222), the Strategic Priority Research Program
of Chinese Academy of Science (Grant NO. XDB34030301). This work is partly
supported by HongLiu Support Funds for Excellent Youth Talents of Lanzhou
University of Technology.

\end{document}